\documentclass[prl,showkeys,twocolumn]{revtex4}
\usepackage{graphicx,amssymb,amsmath}
\usepackage{hyperref}
\usepackage{enumitem}
\usepackage{subfig}
\usepackage{empheq}
\usepackage{color,soul}
\usepackage{mathrsfs}
\begin{document}

\title{The coalescing colony model: mean-field, scaling, and geometry}

\author{Giulia Carra}
\affiliation{Institut de Physique Th\'{e}orique, CEA, CNRS-URA 2306, F-91191,
Gif-sur-Yvette, France}
\author{Kirone Mallick}
\affiliation{Institut de Physique Th\'{e}orique, CEA, CNRS-URA 2306, F-91191,
Gif-sur-Yvette, France}
\author{Marc Barthelemy}
\email{marc.barthelemy@ipht.fr}
\affiliation{Institut de Physique Th\'{e}orique, CEA, CNRS-URA 2306, F-91191,
Gif-sur-Yvette, France}
\affiliation{Centre d'Analyse et de Math\'ematique Sociales, (CNRS/EHESS) 190-198, avenue
de France, 75244 Paris Cedex 13, France}


\begin{abstract}

  We analyze the coalescing model where a `primary' colony grows and
  randomly emits secondary colonies that spread and eventually
  coalesce with it. This model describes population proliferation in
  theoretical ecology, tumor growth and is also of great interest for
  modeling the development of cities.  Assuming the primary colony to
  be always spherical of radius $r(t)$ and the emission rate
  proportional to $r(t)^\theta$ where $\theta>0$, we derive the
  mean-field equations governing the dynamics of the primary colony,
  calculate the scaling exponents versus $\theta$ and compare our
  results with numerical simulations. We then critically test the
  validity of the circular approximation and show that it is sound for
  a constant emission rate ($\theta=0$). However, when the emission
  rate is proportional to the perimeter, the circular approximation
  breaks down and the roughness of the primary colony can not be
  discarded, thus modifying the scaling exponents.

\end{abstract}


\maketitle
%
%
%
%

Dispersal models have been used extensively to investigate the
proliferation of animal colonies in theoretical
ecology~\cite{Shigesada:1997, Clark:2001} and as a simplified model
for the growth of cancerous tumors~\cite{Iwata:2000,
  Haustein:2012}. Such models are also good candidates for describing
the growth of the built-area of cities \cite{Barthelemy:2016} for
which we now have empirical data over long periods of time
\cite{Angel:2005}. The main feature of dispersal models is the
concomitant existence of two growth mechanisms. The first process is
the growth of the main -- so-called primary -- colony, which occurs
via a reaction-diffusion process (as described by a FKK-like equation
\cite{Fisher:1937,Shigesada:2002}) and leads to a constant growth with
velocity $c$, depending on the details of the system. The second
ingredient is {\it random dispersal} from the primary colony, which
represents the emergence of secondary settlements in the framework of
animal ecology, the development of metastatic tumors, or, in the urban
sprawl case, the creation of small towns in the periphery of large
cities. In the real world, dispersion follows privileged directions
under the effect of external forces such as blood vessels, winds and
rivers, or transportation networks for cities but in a first approach,
these anisotropic effects will be neglected.  We will assume that
secondary colonies also grow at the velocity $c$ and will eventually
coalesce with the primary colony, leading to a larger primary colony
whose time-dependent size will depend on the emission rate.

A classical way to study dispersal is through the \textit{dispersal
  kernel} representing the probability distribution of dispersal
distances and various forms  for these
  kernels have been  discussed~\cite{Lewis:2016}.
  A different approach has been introduced by
Kawasaki and Shigesada in~\cite{Shigesada:1997, Shigesada:2002} who
proposed the use of simple models to tackle this challenging
problem. We shall follow this point of view and study 
  the \textit{coalescing
  colony model} where a primary colony
grows at radial velocity $c$ and  emits  a
secondary colony at a rate $\lambda$ 
 and  at a distance $\ell$ from its border (long-range
dispersal). The variable $\ell$ can be drawn  from a
probability distribution $P(\ell)$ but  we consider   here that the
secondary colonies are emitted at a constant distance $\ell_0$ from the
boundary of the primary colony 
 (i.e. $P(\ell) = \delta(\ell - \ell_0)$).
 Besides, we  assume that  each secondary colony  also grows with
the same radial speed $c$ and does not emit tertiary 
 colonies. The dependence of  the emission rate  on the colony size
 is taken into account by the functional form  
\begin{equation}
\lambda(r) = \lambda_0 r^{\theta}~,
\end{equation}
 $r$ being  the radius of the primary colony and 
 $\theta \geq 0$. When $\theta=0$ the growth rate is independent
from the primary colony size,  for $\theta=1$ it is proportional to its
perimeter and for $\theta=2$ to  its area. 

\begin{figure}[h!]
\begin{center}
\includegraphics[angle=0, width=0.3\textwidth]{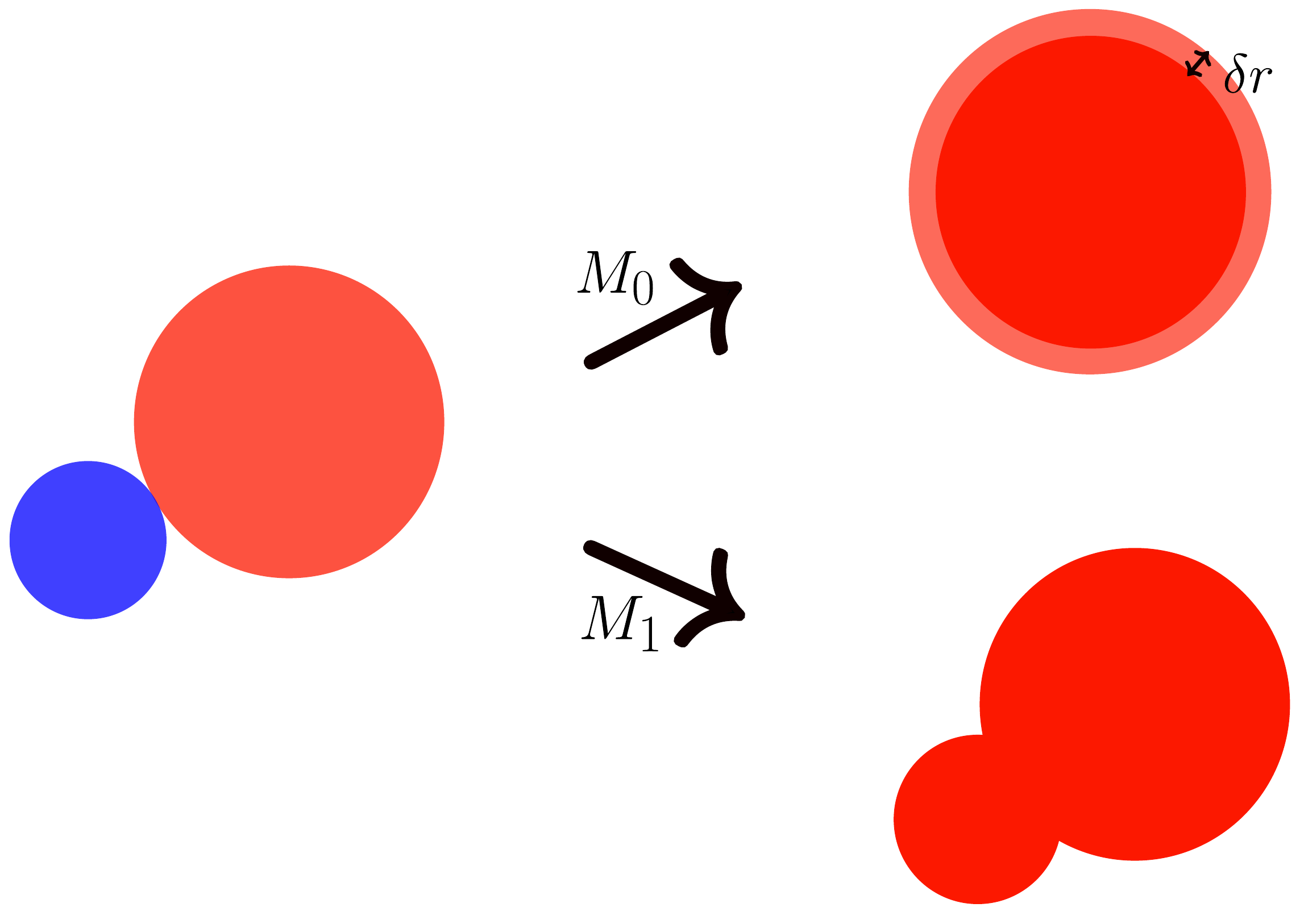} 
\end{center}
\caption{Example of coalescence in  models $M_0$ and $M_1$.
 In $M_0$, the primary colony (in red) remains circular and the area 
 of the secondary colony is  evenly  distributed on the rim; in  $M_1$,
 the shapes are simply  `concatenated'.}
\label{fig:i}
\end{figure}

Coalescence happens when a secondary colony of radius $r_2$ intersects
with the primary one, of radius $r$, and becomes part of the latter.
We shall consider two variants of the process.  In the first version
of the model, denoted by the $M_0$ model, we assume that the primary
colony remains circular after coalescence (see Fig.~\ref{fig:i}), and
has a new radius $r'$ given by
\begin{equation}
{r'}^2=r^2+r_2^2~.
\end{equation}
This interesting model was discussed in \cite{Shigesada:2002} but a
full quantitative understanding of the radius $r(t)$ is still
lacking. Here, we present a microscopic derivation of the dynamics of
the $M_0$ model, in the mean-field approximation, and study its
solutions as a function of the parameter $\theta$. In particular, we
derive a simplified equation that preserves the physics of the system
and allows to extract the scaling behavior for the main quantities of
interest. Our predictions are then tested with numerical simulations.

In the  second part of this work, we discuss the importance of the circular
approximation and its impact on the scaling behaviors. We
introduce a modified version of the process, referred to as  the $M_1$
model, in which  after  coalescence  the secondary
colony merges into the  primary colony and   the shape of the
primary colony does not remain circular. This
important difference between models $M_0$ and $M_1$  is illustrated in the Fig.~\ref{fig:i}.


We now derive the main equations for the model $M_0$. We recall that
$\lambda(t_i)dt_i$ represents the probability to emit a colony in the
interval $[t_i, t_i + dt_i]$ and we denote by ${t_i}'$ the time of
coalescence of a colony emitted at time $t_i$. The condition of coalescence is
given by
\begin{equation}
\label{eq:coal}
r({t_i}') + c {t_i}' = \ell_0 + r(t_i) + c t_i~,
\end{equation}
which defines -- formally -- the function $f$ such that 
\begin{equation}
{t_i}' =  f(t_i)~.
\end{equation}
The mean-field approach that we propose here consists in
neglecting the fluctuations of this function $f(t)$ and to consider
that it is the same for all secondary colonies. The evolution of the area of the primary colony is thus  given by
\begin{equation}
\label{eq:dA}
\frac{dA}{dt} = 2 \pi r c + \int d{t_i} \lambda(t_i) {\delta}(t - f(t_i)) \pi c^2 (t - t_i)^2~,
\end{equation}
where the first term of the rhs is due to short-range dispersion and
the second term represents the coalescence with secondary colonies. This
 leads to 
\begin{equation}
\label{eq:dA_2}
\frac{dA}{dt} = 2 \pi r c + \lambda(f^{-1}(t))\mid \left[f^{-1}(t)\right]' \pi c^2 (t - f^{-1}(t))^2 \mid~.
\end{equation}
We call  $x(t)$  the radius of the colony
absorbed at time $t$, given by $x(t) = c (t -
f^{-1}(t))$. Injecting this quantity in Eqs.~\eqref{eq:coal} and \eqref{eq:dA_2}, we
 obtain  the Kawasaki-Shigesada system of equations
\cite{Shigesada:2002} 
\begin{empheq}[left = \empheqlbrace]{align}
  \frac{dr}{dt} &= c + \frac{\lambda_0 {\left[r\left(t - 
\frac{x(t)}{c}\right)\right]}^{\theta}}{2\pi r(t)} \left(1 - \frac{\dot{x}(t)}{c}\right)\pi x(t)^2  ~,\label{eq:Shi_1}\\
  \ell_0 &= r(t) - r\left(t - \frac{x(t)}{c}\right) +
  x(t)~.\label{eq:Shi_2}
\end{empheq}

In the long time regime,  $t \gg x(t) / c$,  
 the system of Eq.~\eqref{eq:Shi_1} and Eq.~\eqref{eq:Shi_2} takes the
simplified form 
\begin{empheq}[left = \empheqlbrace]{align}
               \frac{dr}{dt} &= c + \frac{\lambda_0 r^{\theta-1} }{2} x(t)^2  ~,\label{eq:Shi_1_s}\\
               x(t) &= \frac{\ell_0}{1 + \frac{\dot{r}}{c}}~.\label{eq:Shi_2_s}
\end{empheq}
These effective equations  allow  us to investigate the behavior of the model
without  altering  the physics of the problem as will be shown by 
comparing the solutions  to numerical simulations.

We first solve the effective system for $\theta = 0$
(ie. $\lambda=\lambda_0$).  Defining $x^{*}$ as the average radius of
a secondary colony just before its coalescence and assuming that it is
constant in time we obtain
\begin{equation}
\label{dr}
\frac{dr}{dt} = c + \frac{\lambda_0}{2r} {x^*}^2~,
\end{equation}
whose solution is
\begin{equation}
r(t) \sim a +  ct + \frac{\lambda_0 {x^*}^2}{2c}\log{\left( \frac{2cr}{\lambda_0 {x^*}^2} + 1 \right)}~.
\end{equation}
 When $t \to \infty$,  the  dominant contribution is 
\begin{equation}
r(t) \sim a + ct + \frac{\mathscr{C}}{c}\log{\left(\frac{c^2
      t}{\mathscr{C}} + 1 \right)}~,
\label{eq:rM0_teta_0}
\end{equation}
with $x^* \simeq \ell_0 / 2$ and
$\mathscr{C} = \frac{\lambda l_0^2}{8}$.  

\begin{figure}[h!]
\begin{center}
\begin{tabular}{cc}
\includegraphics[angle=0, width=0.25\textwidth]{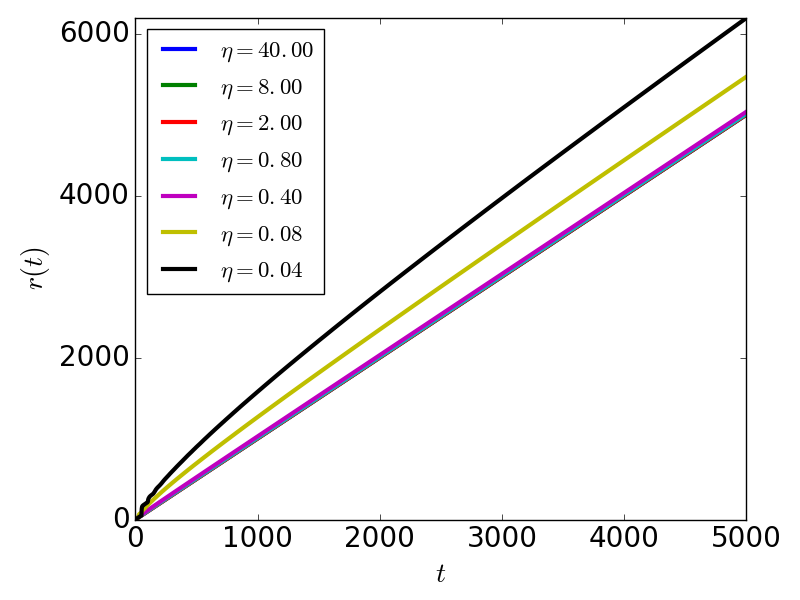}  &
\includegraphics[angle=0, width=0.25\textwidth]{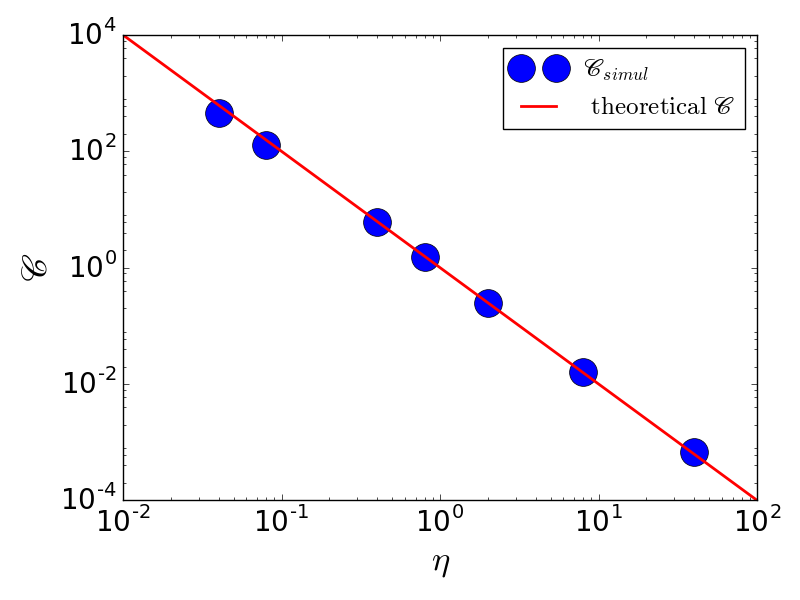}  \\
\includegraphics[angle=0, width=0.25\textwidth]{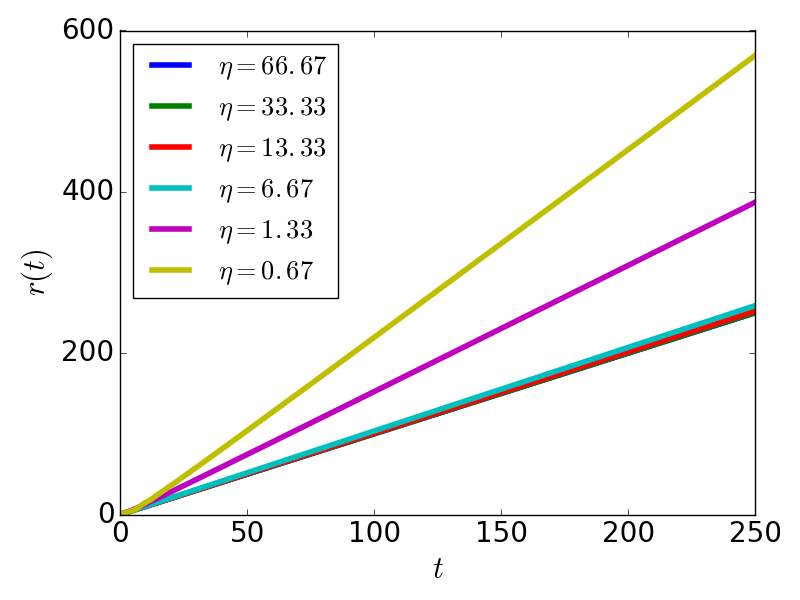}  &
\includegraphics[angle=0, width=0.25\textwidth]{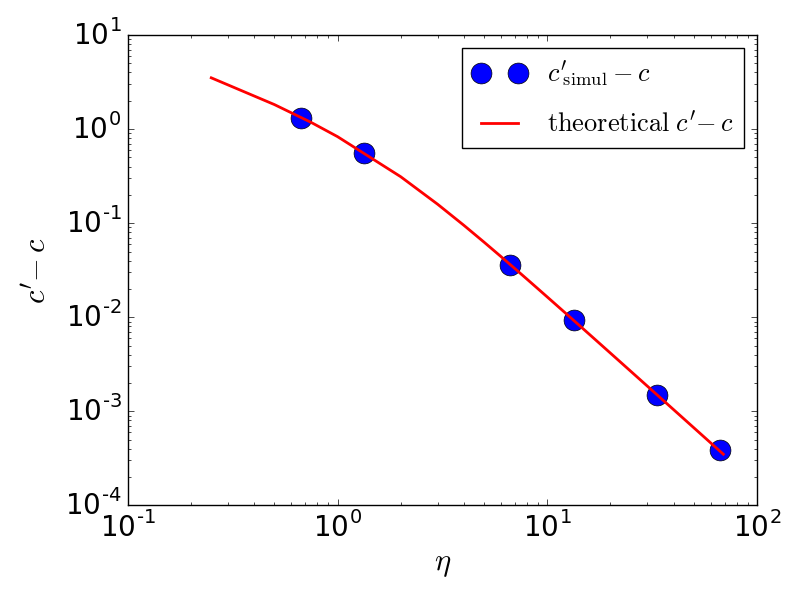} 
\end{tabular}
\end{center}
\caption{(Top) Case $\theta = 0$: (Top-Left) Plot of $r(t)$ versus $t$ for
  different values of the parameter $\eta$, averaged over $10$
  simulations. (Top-Right) Plot of $\mathscr{C}_{simul}$ versus $\eta$
  extracted from the empirical fit. The theoretical prediction is
  shown in red.  (Bottom) Case $\theta = 1$: (Bottom-Left) $r(t)$ versus
  $t$ for different values of the parameter $\eta$, obtained averaging
  over $100$ simulations. (Bottom-Right) $c' - c$ versus $\eta$ as
  obtained from the empirical fit. In red, the theoretical
  prediction (Eq.~\eqref{eq:c_primo}).}
\label{fig:1}
\end{figure}

We perform numerical simulations with a constant $\lambda_0 = 0.5$
and $c = 1$, for different values of the emission distance
$\ell_0$. It is useful 
 to  introduce  $\eta = \frac{2c}{l_0 \lambda_0}$,
which represents the ratio between the emission time
$\tau_e = 1/\lambda_0$ and the coalescence time
$\tau_c = \ell_0 /(2c) $.
In Fig.~\ref{fig:1} Top-Left,  we plot the radius of the
primary colony $r(t)$ versus $t$. We then perform on these data a two
parameters fit with a function of the form
\begin{equation}
g(t) = a + \frac{\mathscr{C}_{simul}}{c}\log{\left( \frac{c^2 t}{\mathscr{C}_{simul}} + 1\right)}~,
\end{equation}
where the  fitting parameters  are $a$ and $\mathscr{C}_{simul}$.
In Fig.~\ref{fig:1} Top-Right we test the results of the fit,
comparing the estimated value $\mathscr{C}_{simul}$ obtained, with its
theoretical value
$\mathscr{C} = \frac{\lambda l_0^2}{8 } = \frac{c^2} {2 \lambda_0
  \eta^2}$. We observe an excellent agreement, showing the validity of
our theoretical calculations.


For  $\theta>1$, we further simplify the system
Eqs.~\eqref{eq:Shi_1_s} and~\eqref{eq:Shi_2_s} by assuming
$\dot{r} \gg c$  and obtain the  single effective 
equation
\begin{equation}
A r(t)^{\theta - 1} \simeq \dot{r(t)}^2 \left( \dot{r(t)} - c \right)
\label{eq:simple}
\end{equation}
with $A = \frac{\lambda_0}{2}c^2 \ell_0^2$. This nonlinear
differential equation captures the physics of the coalescence and
allows us to extract the large-time behavior of the main quantities of
interest in this problem. In particular, assuming scaling laws at
large times of the form $r(t) \sim at^{\beta}$ and $x(t) \sim dt^{-\alpha}$,
Eq.~\ref{eq:simple} yields
\begin{equation}
\beta = \frac{3}{4 -\theta}~, \qquad \qquad  \alpha = \beta - 1~.
\label{eq:beta}
\end{equation}
 Note  that for $\theta \rightarrow 4 $, we have $\beta \rightarrow
 \infty$, the radius grows faster than a power law  and 
explodes exponentially. For $\theta = 1$, we obtain
$\alpha =0, \beta =1$ which means that  we have  $x(t) = x^*$
independent of $t$ and a linear behavior of $r(t)$. From
Eq.~\eqref{eq:Shi_1} we  deduce that the radial velocity
$c'$ is given by 
\begin{equation}
c' = c + \frac{\lambda_0}{2} {x^*}^2
\end{equation}
and the value of $x^*$ can be obtained by solving 
 Eq.~\eqref{eq:Shi_2} that can be written as
\begin{equation}
\label{eq:c_primo}
\frac{\lambda_0 }{2c} {x^*}^3 + 2{x^*} - \ell_0 = 0~.
\end{equation}
This result, for the specific case of $\theta = 1$, was first obtained
by Shigesada and Kawasaki~\cite{Shigesada:1997}. We test this result
numerically (with $\lambda_0 = 0.3$, $c = 1$, and for different values
of the emission distance $\ell_0$) and in Fig.~\ref{fig:1} Bottom-Left
we plot the radius of the primary colony $r(t)$ versus $t$. A linear
fit allows us to obtain an estimate for the radial velocity $c'$ that
we compare in Fig.~\ref{fig:1} Bottom-Right with the theoretical
prediction of Eq.~\eqref{eq:c_primo}. Here also, an excellent
agreement is observed.

\begin{figure}[h!]
\includegraphics[angle=0, width=0.4\textwidth]{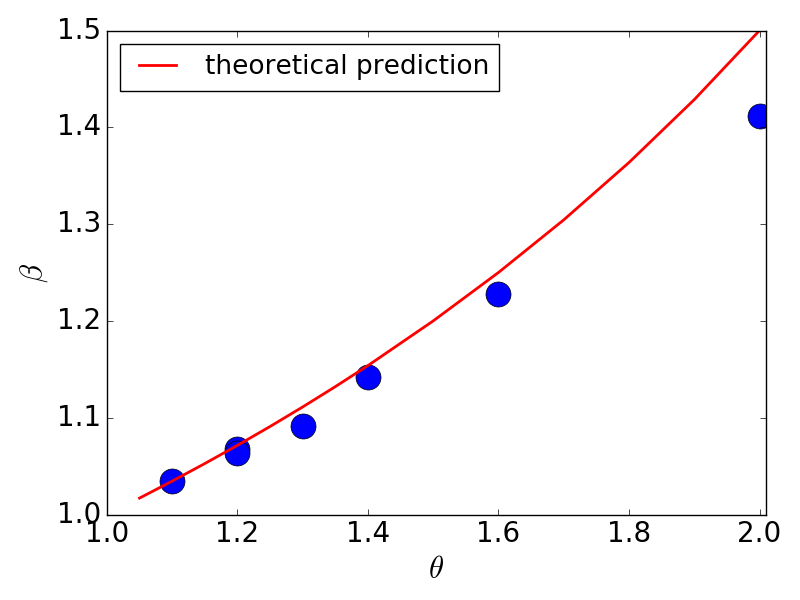} 
\caption{Plot of  the exponent $\beta$ as a function of  $\theta$, obtained from a power
  law fit on $r(t)$ versus $t$ for the values of $\eta^2\lambda /2$
  maximizing the range $[t_{min}, t_{max}]$ in our simulation domain.}
\label{fig:3}
\end{figure}

For $\theta > 1$, the theoretical analysis predicts that the leading
order characterized by a scaling behavior given by
Eqs.~\eqref{eq:beta} can be observed in a range of time $t_{min}\ll t\ll t_{max}$ which 
depend on $\theta$ and on the parameter $\eta^2 \lambda / 2$ (see the
supplementary material for details on this point). We performed numerical simulations for
$\theta = [1.1, 1.2, 1.3, 1.4, 1.6, 2.0]$, with the parameter $c = 1$
and $\lambda = 0.001$ (and additional simulations with the parameter
$\lambda = 0.005$ for $\theta = 1.2$). In Fig.~\ref{fig:3}, we plot the values of the exponents $\beta$
obtained by power law fits and we compare it with the theoretical
prediction (Eq.~\eqref{eq:beta}) in red. We observe a good agreement with some deviations
for higher values of $\theta$ which is probably due the small range
$[t_{min}, t_{max}]$ in this case.

\begin{figure}[h!]
\includegraphics[angle=0, width=0.35\textwidth]{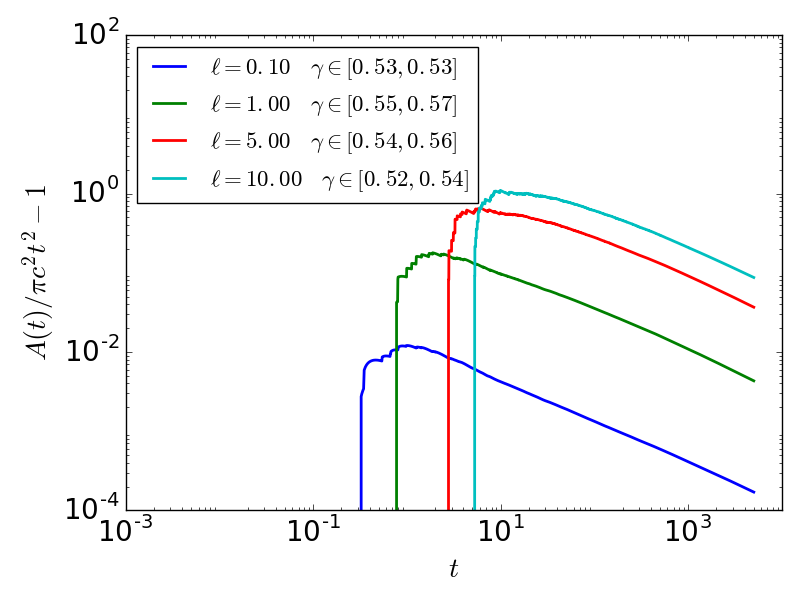}
\includegraphics[angle=0, width=0.35\textwidth]{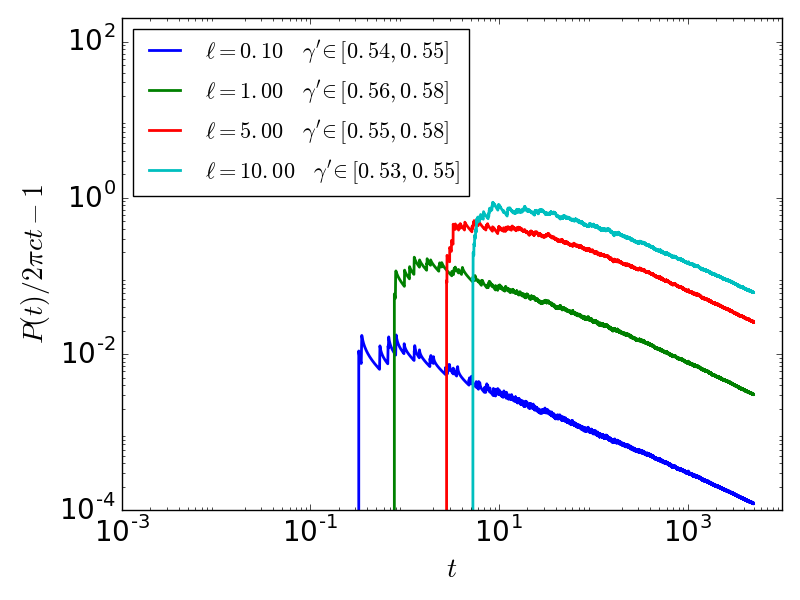} 
\caption{(Left) $A(t)/(\pi c^2t^2) - 1$ versus time $t$. (Right)
  $P(t)/(2\pi ct)- 1$ versus $t$. These results are obtained for
  different values of $\ell$ and are averaged over $100$ numerical
  simulations. For each value of $\ell$, we report in the inset the values of the exponents
  $\gamma$ and $\gamma'$ obtained by fitting these plots.}
\label{fig:2b}
\end{figure}

The Shigesada-Kawasaki coalescing model is based on the circular
approximation. The validity of this hypothesis will be investigated by
simulating the $M_1$ model where we respect the geometry of the
coalescence process.  We first consider a constant emission rate
$\lambda(r) = \lambda_0$ and assume that the area $A$ and the
perimeter $P$ obey to a power law scaling of the form
\begin{equation}
A(t) \sim t^{\mu} \qquad P(t) \sim t^{\nu}~.
\label{eq:beh}
\end{equation}
Performing a power-law fit on the empirical results, we obtain
$\mu  \approx 2$ and $\nu \approx 1$ (see Supplementary Material
for details and plots). These results can be compared with those
 obtained in  the model $M_0$.  In Fig.~\ref{fig:2b},
we plot $A(t)/(\pi c^2t^2) - 1$ and $P(t)/(2\pi ct) - 1$ versus $t$:
these quantities both go to zero for large values of $t$, suggesting
that at large times, the dominant behavior of the $M_0$ and the $M_1$
models are the same with $A(t) \sim \pi c^2t^2$ and
$P(t) \sim 2 \pi c t$. Hence, for $\theta = 0$ and large value of $t$,
the circular approximation appears to be valid.
If  we assume that the  sub-dominant corrections are
described by the  scaling forms
\begin{align}
\frac{A(t)}{\pi c^2 t^2} - 1 \sim t^{-\gamma}\qquad \quad \frac{P(t)}{2\pi c t} - 1 \sim t^{-\gamma'}
\end{align}
the numerical results suggest that $\gamma \sim 0.5$ and
$\gamma' \sim 0.5$ showing that the corrections to the dominant term
are decaying as a power law in model $M_1$, in contrast with the
logarithmic correction observed in the $M_0$ model (see Supplementary Material).

\begin{figure}
\begin{minipage}{.5\linewidth}
\centering
\subfloat[]{\label{main:a}\includegraphics[angle=0, width=1.0\textwidth]{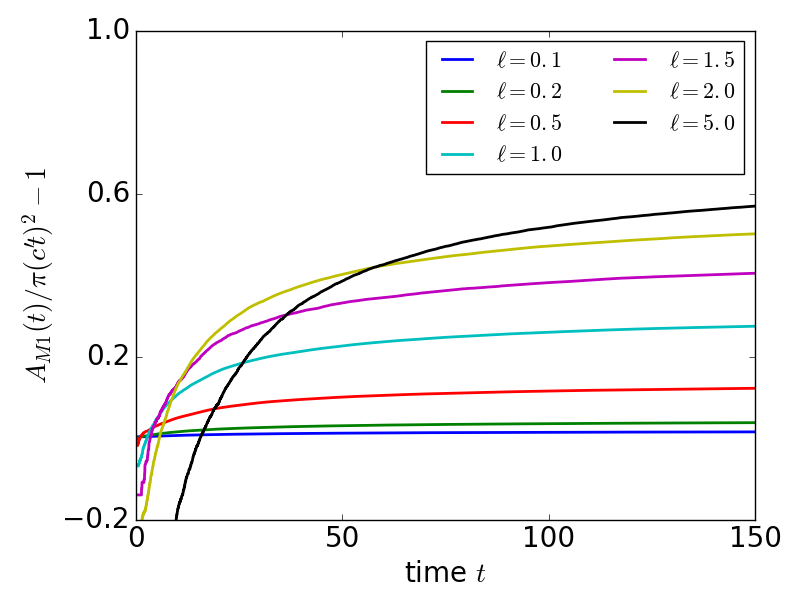}}
\end{minipage}%
\begin{minipage}{.5\linewidth}
\centering
\subfloat[]{\label{main:b}\includegraphics[angle=0, width=1.0\textwidth]{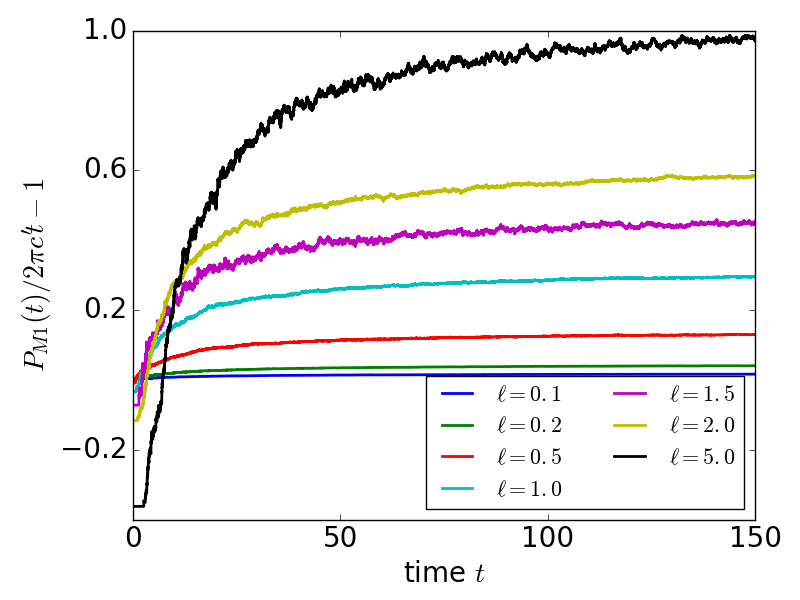}}
\end{minipage}\par\medskip
\centering
\subfloat[]{\label{main:c}\includegraphics[angle=0, width=0.25\textwidth]{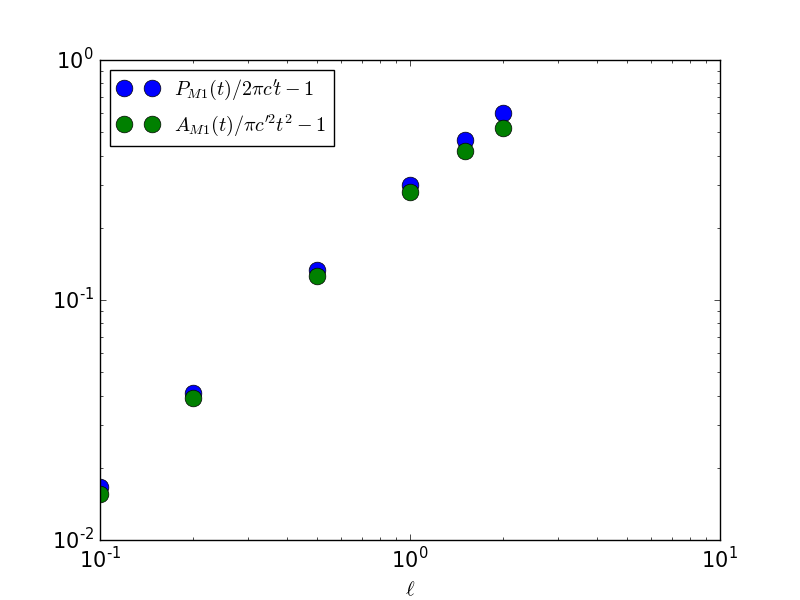}}
\caption{(a) $A(t)/(\pi c^2t^2) - 1$ versus  $t$ for different values of
  $\ell$. (b) $P(t)/(2\pi ct)- 1$ versus  $t$ for different values of
  $\ell$. (c) Plot of $f_1(\ell)$ and $f_2(\ell)$ versus $\ell$ for $t = 200$. These results are
  obtained by averaging over $100$ simulations.}
\label{fig:M1_1_prefactor}
\end{figure}

We now focus on the simulation results obtained for the $M_1$ model
characterized by an emission rate $\lambda$ given by 
\begin{equation}
\lambda(t) = \lambda_0 P(t)~,
\end{equation}
where $P(t)$ is the total perimeter of the primary colony at time $t$,
which corresponds to the case $\theta=1$ in the model $M_0$.  The
simulations results for the area $A(t)$ and the perimeter $P(t)$ of
the primary colony (see Supp. Mat.)  suggest that we still have
$\mu \approx 2$ and $\nu \approx 1$ as in the $M_0$ model. We can go
further and investigate the prefactor. We recall that for the $M_0$
model with $\theta = 1$, the radius of the primary colony increases
with an effective radial velocity $c' > c $. In
Fig.~\ref{fig:M1_1_prefactor} we plot the quantities
$A(t)/\pi c'^2 t^2 - 1 $ and $P(t) / 2 \pi c' t - 1$; if the prefactor
is the same of the $M_0$ model we should find (as we did for
$\theta = 0$) that these quantities tend to zero for large values of
$t$. In Fig.~\ref{fig:M1_1_prefactor}-(a)-(b), we see that these two
quantities tend to a constant that depends on $\ell$. We can therefore
write
\begin{equation}
A(t) = \pi c'^2(1 + f_1(\ell)) t^2 \qquad P(t) = 2\pi c'(1 + f_2(\ell)) t~.
\end{equation}
and we observe numerically that $f_1\equiv f_2$ (see
Fig.~\ref{fig:M1_1_prefactor}-(c)),
 demonstrating  that the circular approximation is not appropriate
for $\theta = 1$.
In order to shed some light on  this behavior, we plot the quantities
$\frac{A(t)}{\pi {\langle r \rangle}^2} -1$ and
$\frac{P(t)}{2\pi {\langle r \rangle}} -1$ where $\langle r \rangle$
is the average radius of the primary colony. 
\begin{figure}
\begin{center}
\begin{tabular}{cc}
\includegraphics[angle=0, width=0.25\textwidth]{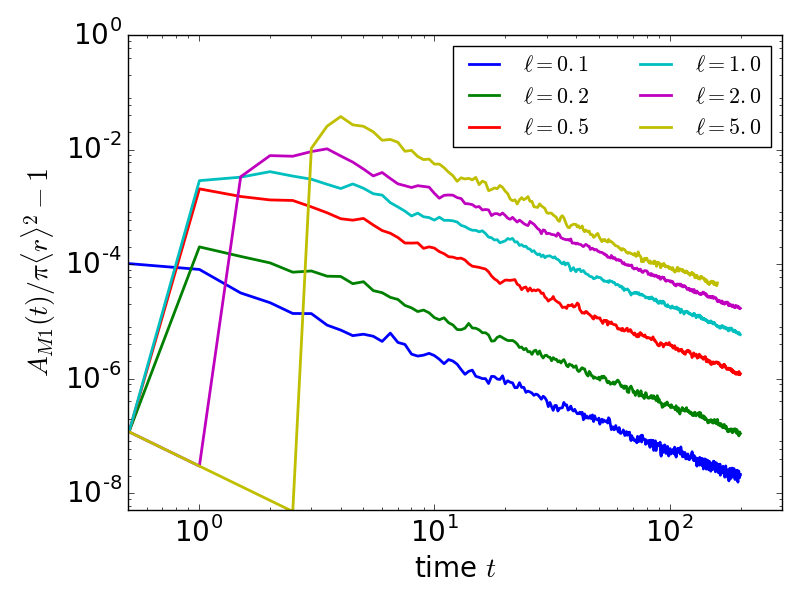} &
\includegraphics[angle=0, width=0.25\textwidth]{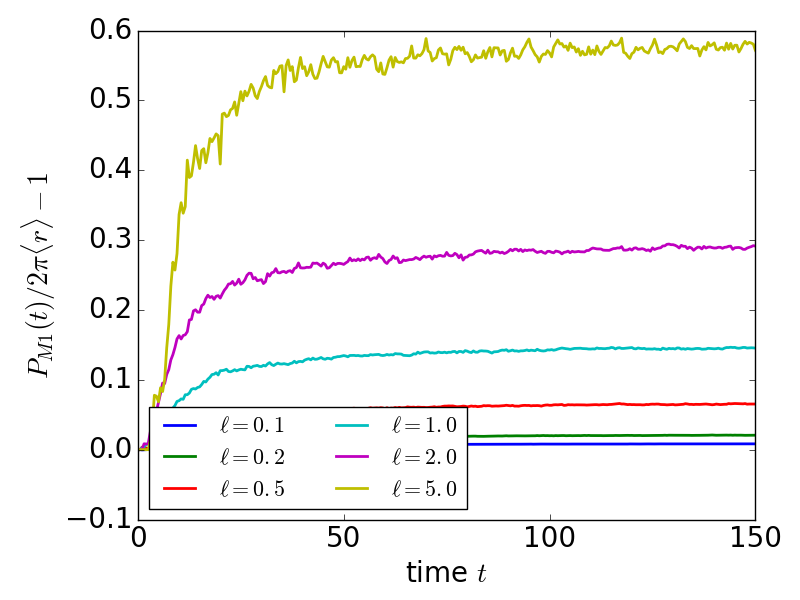}
\end{tabular}
\end{center}
\caption{(Left) $\frac{\pi {\langle r \rangle}^2}{A(t)} -1$ versus $t$ for
  different values of $\ell$. (Right)
  $\frac{P(t)}{2\pi {\langle r \rangle}} -1 $ for different values of
  $\ell$. The results are obtained averaging over $100$ simulations.}
\label{fig:A_cor_2_1}
\end{figure}
The results  shown in Fig.~\ref{fig:A_cor_2_1}  suggest that
the perimeter cannot be described by a  circle, signaling a
breakdown of the circular approximation (even if from the point of
view of the area the system behaves approximately as a circle). 
\begin{figure}[h!]
\begin{center}
\includegraphics[angle=0, width=0.25\textwidth]{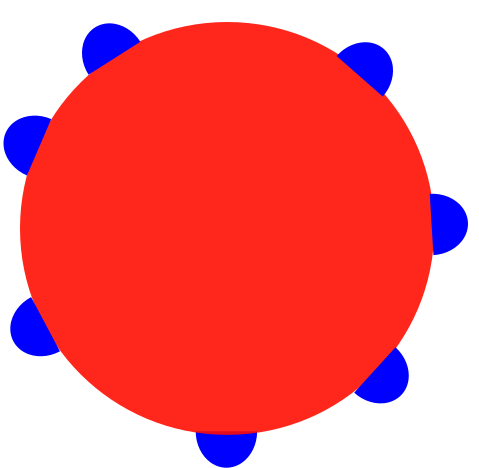} 
\end{center}
\caption{A simplified representation of the primary colony in the
  model $M_1$ with $\theta = 1$ (here we have $n=7$). }
\label{fig:pic_p}
\end{figure}
To visualize the shape of the system, we consider a simplified
picture where the primary colony is described as a circle of radius
$\langle r \rangle$ to which  $n$ semicircles of average
radius $\delta$ are attached (see Fig.~\ref{fig:pic_p} for an illustration). The
maximum number of semicircles is $N = \frac{\pi \langle r
  \rangle}{\delta}$, and we have for this toy model
\begin{equation}
\frac{P}{2\pi \langle r \rangle} - 1 = \frac{n}{N}\left(\frac{\pi}{2} - 1 \right)
\end{equation}
leading to  a value in the range $[0,\frac{\pi}{2} - 1] $, consistent with the
result of Fig.~\ref{fig:A_cor_2_1} (right). This figure also suggests that $n$
increases with the dispersion distance $\ell$, while for small value
of $\ell$ the secondary colonies are quickly absorbed.

\begin{figure}[h!]
\begin{center}
\begin{tabular}{cc}
\includegraphics[angle=0, width=0.25\textwidth]{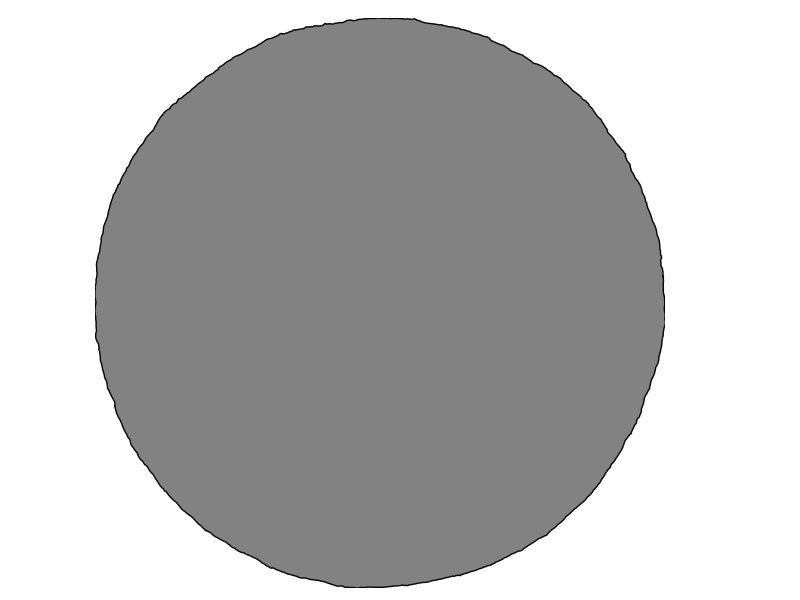} &  
\includegraphics[angle=0, width=0.25\textwidth]{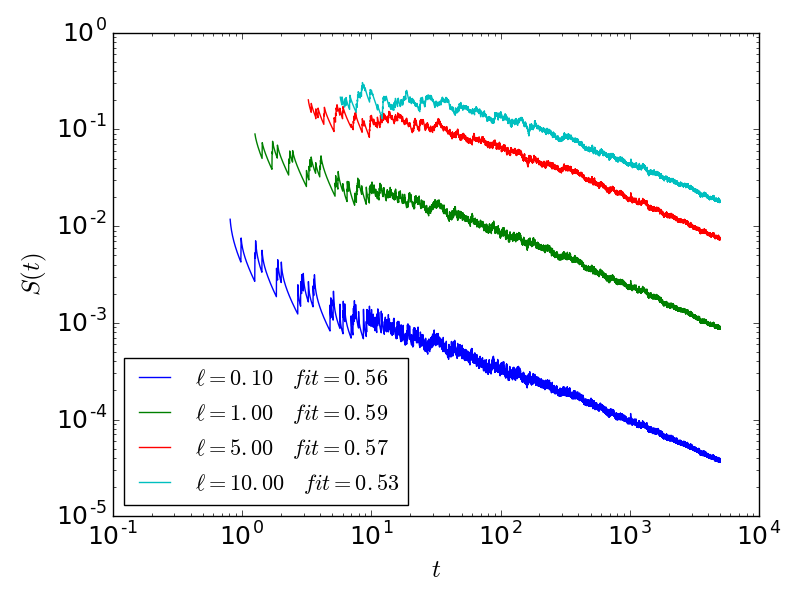}  \\
\includegraphics[angle=0, width=0.25\textwidth]{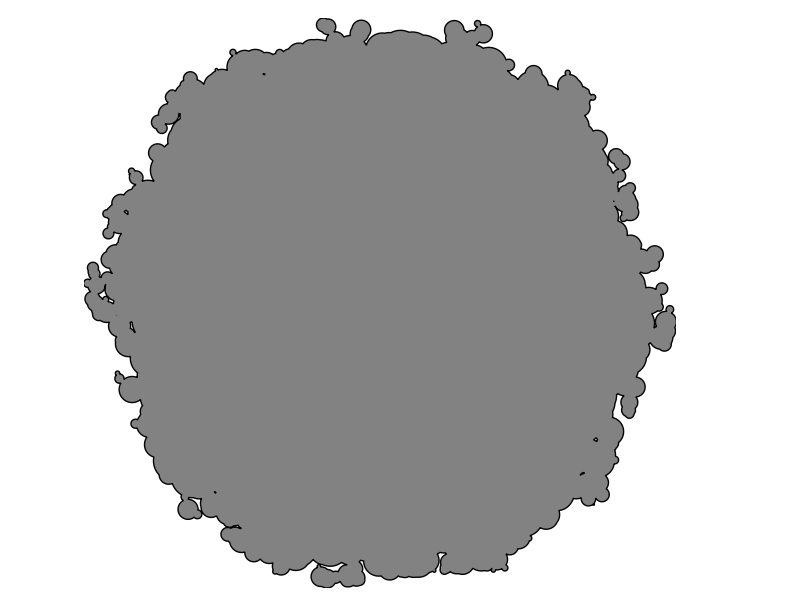} &
\includegraphics[angle=0, width=0.25\textwidth]{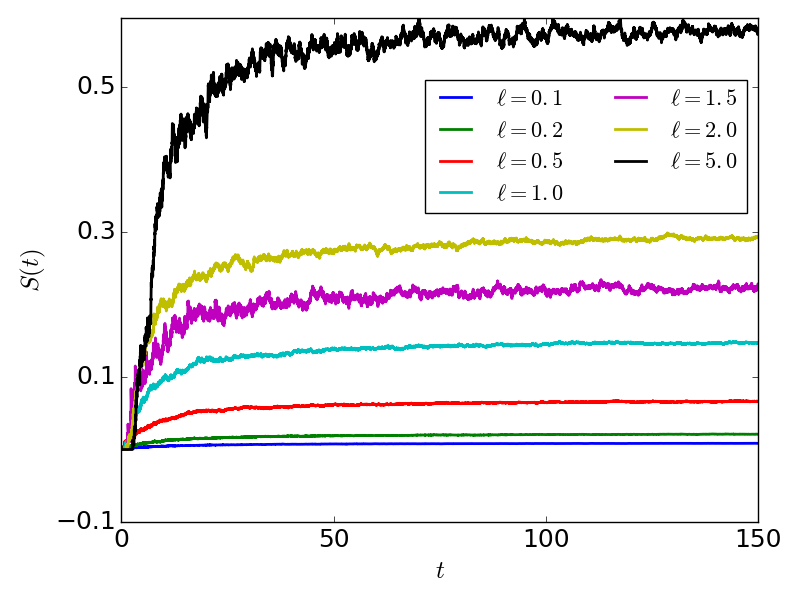}
\end{tabular}
\end{center}
\caption{(Left) Example of shapes obtained for the primary colony in
  the $M_1$ model. (Top-left) Case $\theta = 0$, $\ell =
  10$. (Top-right) Case $\theta = 1$, $\ell = 2$. (right) $S(t)$
  versus $t$. On the top we have the behavior for $\theta = 0$
  averaged over $10$ simulations and on the bottom $\theta = 1$
  averaged over $100$ simulations.  }
\label{fig:2b5}
\end{figure}

The circularity of the primary colony can be probed further
  with the observable 
\begin{equation}
S(t) = P(t) / (2\sqrt{\pi A(t)}) - 1~.
\end{equation}
For a perfect circle $S(t)=0$, whereas $S(t)>0$ estimates the
`rugosity' of the system. The results shown in Fig.~\ref{fig:2b5}(top)
indicate that for $\theta = 0$, $S(t)$ is larger than zero but tends
to zero for large values of $t$ as expected from the previous
discussion and the model $M_0$ seems to be a sound approximation when
$\theta = 0$.  But, for $\theta = 1$, this is not true anymore: we observe
in Fig.~\ref{fig:2b5}(bottom) that $S(t)>1$ and that $S(t)$ tends to a
constant for large $t$, consistently with the previous results.
%
%


We developed the general framework allowing the theoretical discussion
of the growth and coalescence process. We discussed the quantitative
predictions for the simpler model in which the emission rate depend on
the exponent $\theta$, while the distance $\ell$ is constant and the
process is isotropic. However, it is possible to integrate other
specific features such as anisotropy or random emission distances and
to investigate how these latter modify the actual results. Also, the
circular approximation that allows for this analytical approach seems
to be justified in cases where the emission rate grows not too quickly
with the size of the primary colony. Otherwise, it is necessary to
take the geometry of the colony into account, and making the
theoretical extremely challenging. This model is very general and versatile, the results obtained are
potentially useful to gain insights into the understanding of
population proliferation, tumor growth and is also of great interest
for modeling the dynamics of complex systems such as the sprawl of cities.

\paragraph*{Acknowledgments}
GC thanks the Complex Systems Institute in Paris (ISC-PIF) for hosting
her during part of this work and for providing the OpenMole platform.
MB  thanks the city of Paris for its funding `Paris 2030'.

\bibliographystyle{prsty}

\section*{Supplementary material}

\subsection*{The model $M_0$: further investigations}
We discuss for $\theta > 1$ the range of validity of the power-law behavior $r(t) \sim t^{\beta}$~, exploring the second order behavior.\\
Assuming the following form for the evolution of the radius of the primary colony with time,
\begin{equation}
r(t)  \sim  at^{\beta} + bt^{\beta'}~,
\label{eq:eff_Shig}
\end{equation}
we consider the simplified Shigesada-Kawasaki system of equations given by
\begin{empheq}[left = \empheqlbrace]{align}
               \frac{dr}{dt} &= c + \frac{\lambda_0 r^{\theta-1} }{2} x(t)^2  ~,\label{eq:Shi_1_s}\\
               x(t) &= \frac{\ell_0}{1 + \frac{\dot{r}}{c}}~;\label{eq:Shi_2_s}
\end{empheq}
with $A = \frac{\lambda}{2}c^2 \ell_0^2$. After some calculations, a development at the first and second order of Eq.~\eqref{eq:Shi_1_s} bring to the following results
\begin{equation}
\beta' = 1~,
\end{equation}
\begin{equation}
a = \left( \frac{2\beta^3}{{\ell}_0^2 \lambda_0c^2} \right)^{\frac{1}{\theta - 4}}~,
\label{eq:av}
\end{equation} 
and 
\begin{equation}
b = \frac{4 - \theta}{15 - 6 \theta}~.
\end{equation}
Being all the parameters determined we can deduce the value of the time $t_{min} = (b/a)^{1/(\beta - 1)}$ starting from which the second order term begins to be smaller than the first order one. Hence, for $ t \gg t_{min}$ we can write $r(t) \sim a t^{\beta}$, neglecting the second order term. After some calculation one gets
\begin{equation}
t_{min} = f(\theta) \left( \frac{2c^{2 - \theta}}{{\lambda}_0 {\ell_0}^2} \right)^{\frac{1}{\theta - 1}}
\label{eq:t_c}
\end{equation}
with
\begin{equation}
f(\theta) = \left( \frac{4 - \theta}{ 15 - 6 \theta}  \right)^{\frac{4 - \theta}{\theta - 1}} \left( \frac{27}{(4 - \theta)^3}  \right)^{\frac{1}{\theta - 1}}~.
\end{equation}
We remark moreover that the Shigesada-Kawasaki equations are valid only if the coalescence of a colony does not cause the coalescence of another secondary colony. This means that the increasing in the radius at time $t$, $\delta r(t)$ has to be smaller than the distance between two successively emitted secondary colonies. The following relation has to be verified
\begin{equation}
\sqrt{r^2 + {x^{*}}^2} - r < \frac{2c}{\lambda_0 r^{\theta}}~.
\end{equation}
From Eq.~\eqref{eq:Shi_2_s} one can write
\begin{equation}
x(t) \simeq \frac{c \ell_0}{\dot{r(t)}}~,
\end{equation}
this implies the following relation
\begin{equation}
x(t) \sim d t^{-\alpha}
\end{equation}
with $\alpha = \beta - 1$ and $d = c \ell_0 / (\beta a)$.
After some calculations one can show that the Shigesada-Kawasaki system of equations is valid only for $t< t_{max}$, with
\begin{equation}
t_{max} = g(\theta)   \left( \frac{2c^{2 - \theta}}{{\lambda}_0 {\ell_0}^2} \right)^{\frac{1}{\theta - 1}}
\label{eq:t_max}
\end{equation} 
and
\begin{equation}
g(\theta) = 2^{\frac{4-\theta}{\theta-1}}\beta~.
\end{equation}	

\begin{figure}[h!]
\begin{center}
\includegraphics[angle=0, width=0.5\textwidth]{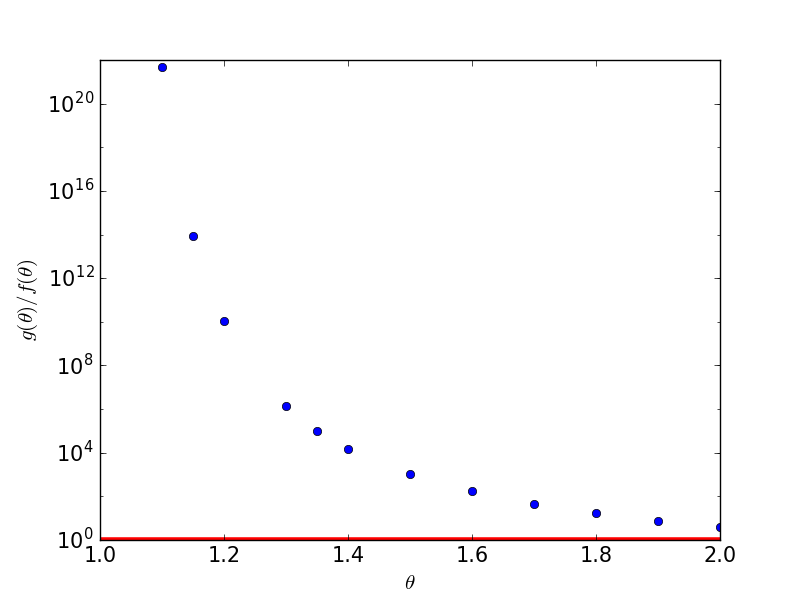} 
\end{center}
\caption{\textbf{$g(\theta)/f(\theta)$ versus $\theta$.}}
\label{fig:m1}
\end{figure}

To summarize, we are able to observe the power-law behavior given by
$r(t) \sim t^{\beta}$ in the range of time for which
$ t_{min} \ll t < t_{max} $. The size of the range of validity depends
on the ratio between $g(\theta)$ and $f(\theta)$. This ratio decreases
when $\theta$ increases as shown in Fig.~\ref{fig:m1}.

In the table~\ref{t_0} we report for different $\theta$, the values of the variable $\eta^2 \lambda_0 / 2$ for which we performed numerical simulations and fit, and the corresponding values of $t_{min}$, $t_{max}$.
\begin{table}[!h]
\begin{center}
    \begin{tabular}{ | l | l |l |l| l|l |l|}
    \hline
     & $\theta = 1.1$  & $\theta = 1.2$ & $\theta = 1.3$ & $\theta= 1.4 $& $\theta = 1.6 $ & $\theta = 2.0$   \\ \hline \hline
     	$\frac{\eta^2 \lambda_0}{ 2}$ &  $0.5$ & $0.8$ & $2.2 $ &$2.2$ & $ 5 $ & $20 $ \\ \hline
     $t_{min}$ & $3 \times 10^{-16}$ & $ 5 \times 10^{-7} $ &  $ 6 \times 10^{-3} $  &$ 5 \times 10^{-2} $ & $ 1.74 $ & $ 30 $    \\ \hline
     $t_{max}$ & $ 1.5 \times 10^6$ & $ 5.7 \times 10^3 $ & $8.1 \times 10^3 $ &$7.7 \times 10^2$ & $ 2.9 \times 10^2 $ & $1.2 \times 10^2 $ \\ \hline

    \end{tabular}
    
\end{center}
\caption{In the table we report for the different $\theta$, the values of the variable $\eta^2 \lambda_0 / 2$ for which we performed the fit, and the corresponding values of $t_c$, $t_{max}$.  }
 	\label{t_0}
\end{table}

\paragraph{Avalanche effect}

We have just discussed that it exists a time $t_{max}$ over which the Shigesada-Kawasaki equations are not valid anymore. Indeed, for $t > t_{max}$ avalanche effects arise. 
\begin{figure}[h!]
\centering
\label{main:a}\includegraphics[angle=0, width=0.5\textwidth]{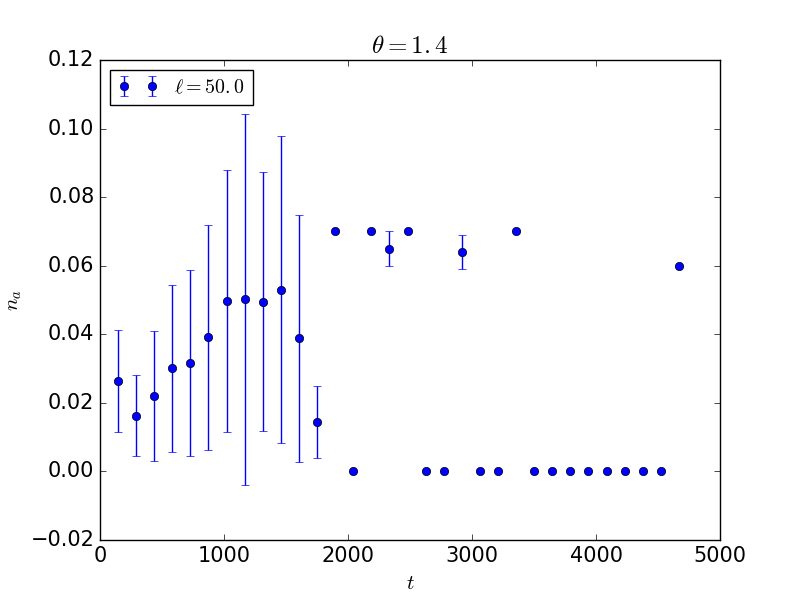}
\label{main:b}\includegraphics[angle=0,width=0.5\textwidth]{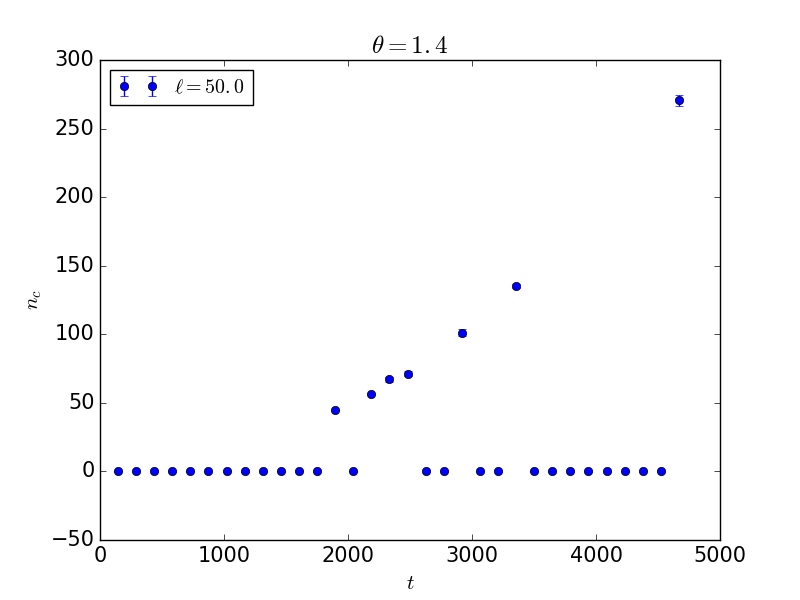}
\label{main:c}\includegraphics[angle=0, width=0.5\textwidth]{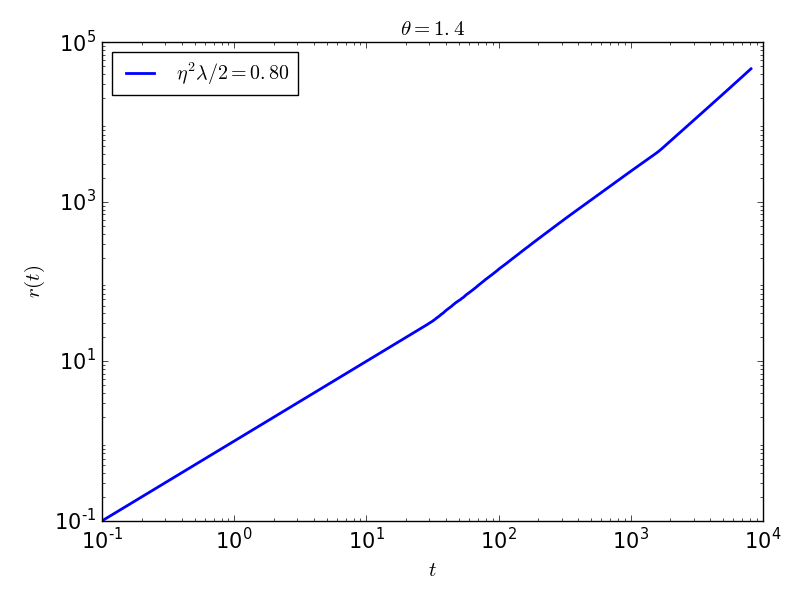}
\caption{(Top) $n_a$ vs. $t$ (Middle) $n_c$ vs. $t$. (Bottom) $r(t)$ vs. $t$. The
  results are obtained averaging over $100$ simulations. For
  $\theta = 1.4$, $\ell = 50$, $c = 1$, $\lambda = 0.001$.}
\label{fig:aval}
\end{figure}
This is due to the high emission rate and means that not only we can have multiple coalescences, (that is more colonies absorbed in a single time step), but the increase in the radius produced by these coalescences can bring to other coalescences before moving to the next time step. Every time this happens we say that we observe an avalanche. In this situation the Shigesada-Kawasaki equations do not held and another treatment of the problem is necessary. This goes beyond the aim of this paper, however we performed numerical simulations to highlight this phenomenon, with the choice of $\theta = 1.4$. At each time step $dt = 0.001$ we count the number of avalanches $n_a$ and the number of total coalescences $n_c$ happened during $dt$ as consequence of the different avalanches.\\
The plots are shown in Fig.~\ref{fig:aval}(a-b) where we observe that at a given time around $t \approx 2000$, the avalanche phenomenon change behavior acquiring more relevance, and bringing to a change in the slope characterizing the behavior of $r(t)$ with time (see Fig.~\ref{fig:aval}-c).

\subsection*{The $M_1$ model: further empirical results}

\subsubsection*{Case $\theta = 0$}

We present here further simulation results obtained for the $M_1$ model with a constant emission rate $\lambda(r) = \lambda_0$.
\begin{figure}[h!]
\begin{center}
\includegraphics[angle=0, width=0.45\textwidth]{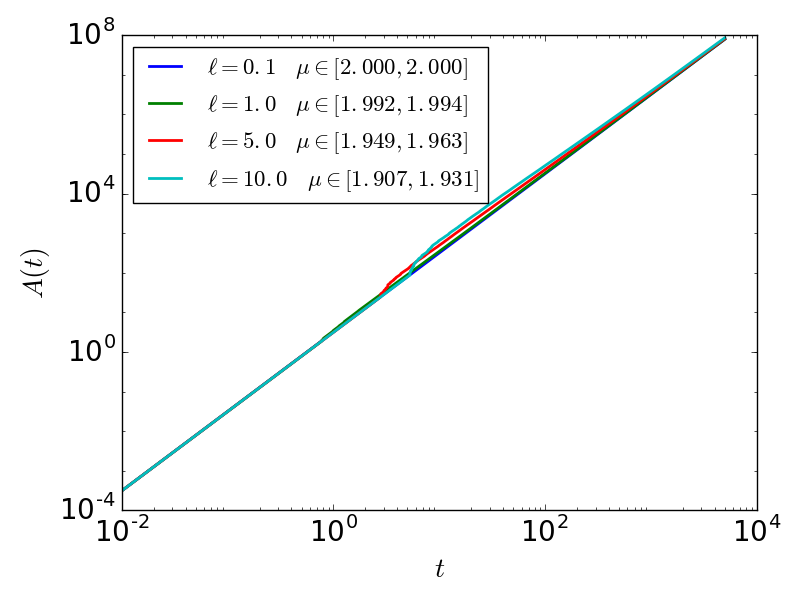} 
\includegraphics[angle=0, width=0.45\textwidth]{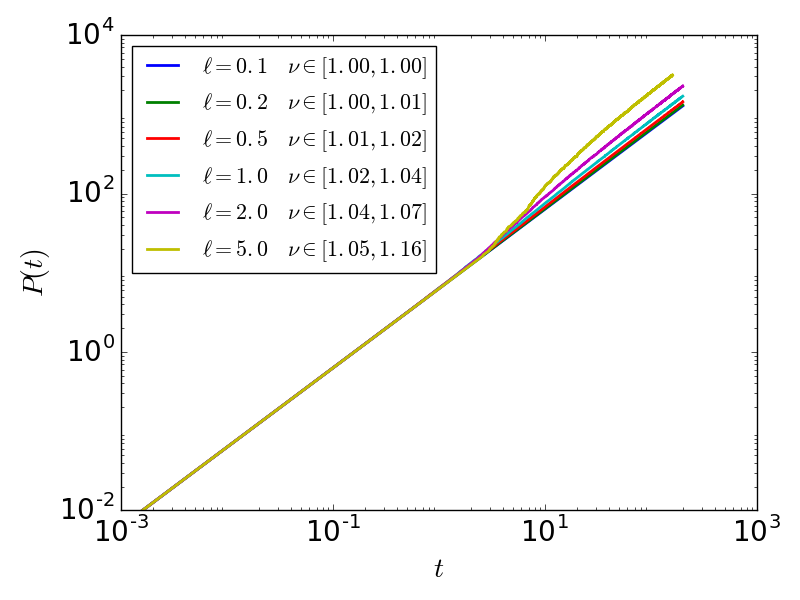} 
\end{center}
\caption{(Top) $A(t)$ vs. $t$. (Bottom) $P(t)$ vs. $t$.  For different
  values of $\ell$ we plot in a log-log scale, the area and the
  perimeter of the primary colony versus time averaged over $10$
  simulations. We perform a power-law fit and the values of the
  exponents obtained are shown in the insets.}
\label{fig:2b2}
\end{figure}
We assume that the area $A$ and the perimeter $P$ of the primary colony obey to a power-law scaling of the form
\begin{equation}
A(t) \sim t^{\mu} \qquad P(t) \sim t^{\nu}~.
\label{eq:beh}
\end{equation}
In Fig.~\ref{fig:2b2} we perform a power-law fit on the empirical
results for two different time regimes. This allows us to examine
eventual finite-size effects: we choose $t > t_{fc}$ and then $t> 100
t_{fc}$, with $t_{fc}$ being the time at which the first coalescence
happens. The values of the exponents $\mu$ and $\nu$ are shown in the
insets of Fig.~\ref{fig:2b2}, with the higher value corresponding to
the choice $t> 100 t_{fc}$. 

We assume that the  sub-dominant corrections are described by the scaling forms
\begin{align}
\frac{A(t)}{\pi c^2 t^2} - 1 \sim t^{-\gamma}\qquad \quad \frac{P(t)}{2\pi c t} - 1 \sim t^{-\gamma'}~.
\end{align}
In the Table~\ref{t_1} we report for the different choice of $\ell$,
the values of $t^*$ and of the exponents $\gamma$ and $\gamma'$. For
each value of $\ell$ the smaller value of the exponent correspond to
the fit for $t> t^*$ and the larger value to the fit range
$t> 10 t^*$.
\begin{table}[!h]
\begin{center}
    \begin{tabular}{ | l | l |l |l| l|}
    \hline
     & $\ell = 0.10$  & $\ell = 1.0$ & $\ell = 5.0$ & $\ell = 10.0 $    \\ \hline \hline
     $\gamma$ & $0.53 - 0.53$ & $ 0.55 - 0.57$ & $0.54 - 0.56$ & $0.52 - 0.54 $   \\ \hline
     $\gamma'$ & $ 0.54 - 0.55$ & $ 0.56 - 0.58 $ & $0.55 - 0.58$ & $0.53 - 0.55$  \\ \hline
    $t^*$ &  $10$ & $30$ & $70$ & $100$  \\ \hline

    \end{tabular}
    
\end{center}
\caption{In the table we report for the different choice of $\ell$, the values of $t^*$ and of the exponents $\gamma$ and $\gamma'$. For each value of $\ell$ the smaller value of the exponent correspond to the fit for $t> t^*$ and the larger value to the fit range $t> 10 t^*$. }
 	\label{t_1}
\end{table}

\subsubsection*{Case $\theta = 1$}

\begin{figure}[h!]
\begin{center}
\includegraphics[angle=0, width=0.45\textwidth]{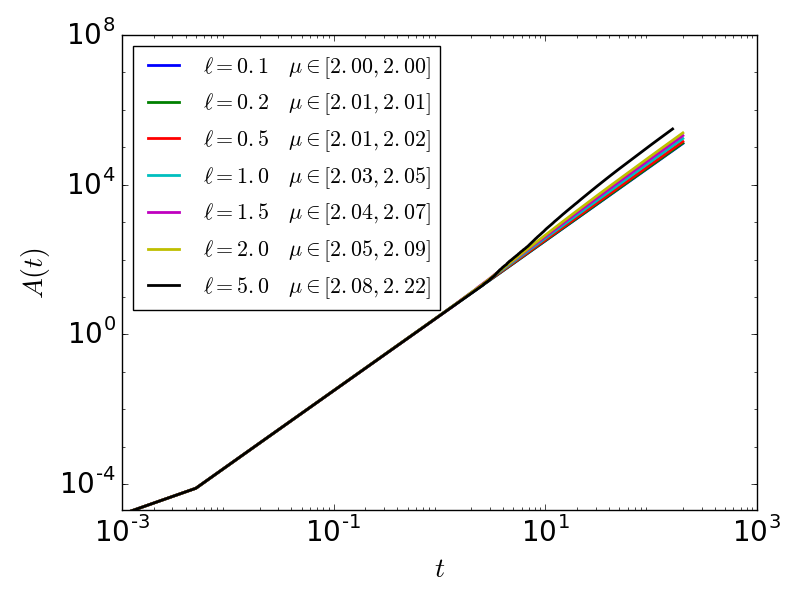} 
\includegraphics[angle=0, width=0.45\textwidth]{./P_versus_t_fit_log_log.png} 
\end{center}
\caption{(Top) $A(t)$ vs. $t$. (Bottom) $P(t)$ vs. $t$.  For different
  values of $\ell$ we plot in a log-log scale, the area and the
  perimeter of the primary colony versus time averaged over $100$
  simulations. We perform a power-law fit and the values of the
  exponents obtained are shown in the insets.}
\label{fig:5}
\end{figure}

The simulations results for the area $A(t)$ and the perimeter $P(t)$
of the primary colony, obtained for different values of $\ell$ are
shown in Fig.~\ref{fig:5}. We assume the power-law behaviors given by
Eq.~\eqref{eq:beh} and we perform a fit on the empirical data. The
values of the exponents obtained (for the time range $t > t_{fc}$ and
$t> 20 t_{fc}$, with $t_{fc}$ the time at which the first coalescence
happens) are shown in the insets of Fig.~\ref{fig:5}. The higher
values correspond to the choice $t> 20 t_{fc}$ and this suggests that,
taking possible finite-size effects into account, one can write
\begin{equation}
\mu \approx 2 \qquad \nu \approx 1~.
\end{equation}

\end{document}